
\input phyzzx
\hoffset=0.375in
\def\P{\rm Panagia et al.\ }
\def\lmc{{\rm LMC}}
\def\rise{{*}}
\def\kpc{\rm kpc}

\def\dlmc{{D_{\rm LMC}}}
\def\dsn{{D_{\rm SN}}}
\def\sn{{\rm SN}}
\def\htp{{\hskip-2pt}}

\centerline{}
\bigskip
\bigskip
\singlespace
\font\bigfont=cmr17
\centerline{\bigfont The Supernova Ring Revisited II:}
\smallskip
\centerline{\bigfont Distance to the LMC}
\bigskip
\centerline{\bf Andrew Gould}
\bigskip
\centerline{Dept of Astronomy, Ohio State University, Columbus, OH 43210}
\smallskip
\centerline{E-mail:  gould@payne.mps.ohio-state.edu}
\bigskip
\vskip 1.5in
\singlespace
\centerline{\bf ABSTRACT}

	I derive an upper limit to the distance to SN 1987A of
$\dsn<46.77\pm 0.76\,\kpc$,
or a distance modulus $\mu_\sn<18.350\pm 0.035$.  If SN 1987A lies
in the plane of the Large Magellanic Cloud (LMC) and hence 500 pc in front of
the LMC center, this implies $\mu_\lmc<18.37\pm 0.04$.  The determination
rests on a new measurement of the caustics in the ionized-emission light
curves of $t_-=75.0\pm 2.6\,$days and $t_+=390.0\pm 1.8\,$days.  The estimate
assumes that the ring is circular.  This assumption is shown to be
consistent with the available data and can be subjected to a future precise
test.  Moreover, I have previously shown that modest deviations from
circularity (eccentricities $e<0.4$) affect the result by $<1\%$.
The upper limit to the distance modulus to M31 is $\mu_{\rm M31} <
24.43\pm 0.06$.  If the response function of the supernova-ring gas
is prompt, then these upper limits become equalities.

Subject Headings:  astrometry, distance scale, Magellanic Clouds, M31
\vskip 0.3in
\centerline{submitted to {\it The Astrophysical Journal}: July 14, 1994}
\bigskip
\centerline{Preprint: OSU-TA-11/94}
\endpage
\normalspace

\chapter{Introduction}

	\P (1991) derived a distance to SN 1987A in the Large Magellanic
Cloud (LMC) from the light curve of the ring illuminated by the supernova.
They obtained $\dsn=51.2\pm 3.1\kpc$ based on their
measurements of the light-curve caustics (i.e., the light travel delay times
to the near and far sides of the ring) at $t_-=83\pm 6\,$days and
$t_+=413\pm 24\,$days and on a previous measurement by Jakobsen et al.\
(1991) of the angular size of the ring axes,
$\theta_-=1.\htp ''21\pm 0.\htp ''03$ and
$\theta_+=1.\htp ''66\pm 0.\htp ''03$.  If the ring is assumed to be
intrinsically circular, these two sets of measurements lead to estimates
for the ring's angle of inclination of
$i_t=41^\circ\htp .7\pm 2^\circ\htp .0$ and
$i_\theta=43^\circ\htp .2\pm 1^\circ\htp .8$ from $t_\pm$ and $\theta_\pm$
respectively.  The fact that these
two estimates for $i$ are consistent with one another
together with what \P considered the
implausibility of a non-circular geometry, convinced them that the ring
should be treated as circular.

	In Paper I (Gould 1994a), I addressed several points.  First,
I developed a method for incorporating all the data on a uniform basis.
Second, I determined how the distance determination would be affected
if the ring were intrinsically elliptical.  I showed that if $i_t$ and
$i_\theta$ are consistent within errors, then the effect of
a small eccentricity, $e$, on the distance estimate is quite small,
$\dsn \rightarrow (1-0.4e^4)\dsn$.  Finally, I noted that \P had
probably mis-measured the position of the light caustics.  I pointed
out that the N III] curve shows a clear peak at 390 days, and N IV] shows
a somewhat messier peak at the same time.
 I suggested that if one used the proper reflection function for a ring
(as described by Dwek \& Felten 1992), then $t_+$ would probably come
out closer to this value.  Because of technical difficulties in
refitting the data (see Appendix), I used the \P values for $t_\pm$
when applying my new formalism.  I deferred a redetermination of $t_\pm$
to the present work (Paper II).  In Paper III, I will discuss the
implications of this new determination for cosmology.

\chapter{Analysis}

	The light curves arise from fluorescence of ionized gas.  The initial
UV flash ionized and heated the gas which is then excited by thermal collisions
and radiates at characteristic wavelengths.  I therefore represent the time
dependence of the radiation of each {\it localized} lump of gas in the
ring as 1) 0 until the UV flash hits the lump, 2) rising linearly to a peak
intensity, $A$, over a time period, $t_\rise$, and finally 3)
exponentially decaying with time scale $\tau$ as the gas cools.  That is,
the fluorescence function, $G(t)$, is given by
$$ G(t) = 0,\ (t<0);\quad G(t) = {A t\over t_\rise},\ (0<t<t_\rise);\quad
G(t) = A\exp\biggl(-{t-t_\rise\over \tau}\biggr),\ (t>t_\rise).\eqn\fluor$$
To find the light curve due to the whole ring, this function must be
convolved with the reflection function, $R(t)$, which characterizes the
light travel delay times to the circumference of the ring.
Dwek \& Felten (1992) have shown that $R$ is given by
$$ R[x(t)] = {\Theta(1-x)\Theta(1+x)\over \pi\sqrt{1-x^2}},\qquad
x(t)\equiv {2t - t_+ - t_-\over t_+ - t_-},\eqn\dwek$$
where $\Theta$ is the Heaviside step function.  The light curve is then
given by the convolution
$$F(t;A,t_\rise,\tau,t_+,t_-) = R(t;t_+,t_-)\otimes G(t;A,t_\rise,\tau).
\eqn\convolv$$
The principal difference between my approach and that adopted
by \P (1991) is that they used a reflection function $R(x) =
\Theta(1-x)\Theta(1+x)/2$.

	To measure $t_-$ and $t_+$, I proceed as follows.  I
measure the data points for the 3 nitrogen ions, N III] 1750$\rm \AA$,
N IV] 1486$\rm \AA$, and N V 1240$\rm \AA$.  I did not attempt to measure
the C III] line for two reasons.  First, as I will show below, the
nitrogen light curves show signs of being inter-related, so that an accurate
estimate of $t_-$ can be obtained only by considering them together.  With
only one carbon line available, there would be no opportunity to check
for such a mutual dependence in carbon.  Second, as also discussed below,
the data for each of the four ions as obtained from two different sources
show significant differences.  However, for C III], the differences are so
severe that the data appear to be unusable unless the source of the
discrepancy is tracked down.

	Next, I assume a ``prompt fluorescent response'' for the ionized gas
and set $t_\rise = 1\,$day.  The UV flash certainly peaked in
the first day or two.  The rise time could be longer than this if, for example,
the ring were optically thick in a given line, or if the population of
a given ion increased with time due to recombination.  My general approach
will be to ignore such effects in the initial analysis where I derive a
relatively precise distance to the LMC within the context of this perhaps
overly simplified model.  Later I will show that any delayed fluorescent
response causes one to overestimate the distance to the LMC.  To measure
the distance to the LMC rather than simply to obtain an upper limit
probably requires a much more detailed model of the fluorescent response
functions such as is presently being developed by Lundqvist (1994).

I note that
since N III] and N IV] are semi-forbidden, they are certainly optically thin.
N V could in principal
be optically thick in which case one might want to model its light curve
differently.
However, it turns out that whether N V is included in or excluded from the
analysis does not affect the results presented here.
I therefore include N V for completeness and treat it as though it were
optically thin.

	I ignore the finite thickness of the ring.  The ring could plausibly
have a thickness of 10 or 20 light-days, but treating it as infinitely thin
is still appropriate.  The reason is that a ring of finite thickness
can be mimicked by incorporating a response function $G$ with a rise time
equal to the effective light crossing time of the ring.  Numerical experiments
show that this light curve looks almost exactly the same as the light curve
from an infinitely thin ring whose circumference lies in the center of the
thick ring.  Since the center of the thick ring is what is observed to
determine the ring angular diameters, $\theta_-$ and $\theta_+$, a
thick-ring model and a thin-ring model lead to nearly identical results.

	Next, I make allowance for the systematic differences
between the intensity measurements as recorded by the GSFC station
(circles in the \P 1991 figures and in my figures below) and the VILSPA
station (triangles in \P and crosses below).  The VILSPA data are
systematically higher than the GSFC data for all three ions.  I therefore
multiply the VILSPA data by a factor, $Q$ which is constant for each ion.

	I then let $A,\tau,Q,t_-$, and $t_+$ all vary to find the
least-squares best fit for each ion.  For each ion, I set the `error' $\sigma$
equal to a constant value such that the $\chi^2$ per degree of freedom
is unity.  I find
$\sigma_{\rm III}\sim\sigma_{\rm IV}\sim 3$ and $\sigma_{\rm V}\sim 5$.
Here and later, intensity is expressed in units of $10^{-14}\rm erg\ cm^{-2}\
s^{-1}$.  Finally, I allow $A,\tau,$ and $Q$ to vary in order to
minimize $\chi^2$ for each value of $t_-$ and $t_+$.

\chapter{Measurement of the Caustic Times $t_\pm$}

\FIG\one{Goodness of fit ($\chi^2$ relative to its minimum value)
for the light curves of N III] ({\it solid}),
N IV] ({\it dashes}), and N V ({\it dot dashes}) as a function of the
time of the first caustic, $x_-$.  The sum of the three curves is shown
as a bold line.  $x_+$ is held fixed at 390 days, but the
curves are insensitive to changes in $x_+$.
}
\FIG\two{Goodness of fit ($\chi^2$ relative to its minimum value)
for the light curves of N III] ({\it solid}),
N IV] ({\it dashes}), and N V ({\it dot dashes}) as a function of the
time of the second caustic, $x_+$.  The sum of the three curves is shown
as a bold line.  $x_-$ is held fixed at 75 days, but the
curves are insensitive to changes in $x_-$.
}
	For all three nitrogen ions, the best fit value for $t_+$ is near
390 days.  The second caustic is therefore well determined.  See Figure \two.
Note that the $\chi^2$ for both the individual ions and their sum deviate
significantly from a parabolic shape.  I therefore estimate the best fit
and error from the ``$2\,\sigma$ interval''.  That is, I find the center
and 1/4 of the length of the range of $t_+$ with $\Delta \chi^2<4$.  I find
$t_+=390.0\pm 1.8$ days.
\FIG\three{Light curve of emission in N III] from the SN 1987A ring.
Intensity is in units of
$10^{-14} \rm erg\ cm^{-2}\ s^{-1}$.  Shown are data points from GSFC
({\it circles}) and VILSPA ({\it crosses}) taken from \P (1991).  See
Appendix below.  The curve is the best fit to the data to eq.\ \convolv\ with
$x_-=75.0\,$days, $x_+=390.0\,$days adopted from best fit to all three ions.
The remaining parameters from eq.\ \convolv\ are $A=65$, $t_\rise=1\,$day, and
$\tau=287\,$days.  VILSPA data have been multiplied by an amplitude correction
factor $Q=0.88$ which minimizes scatter of the fit.  $Q$ is determined with
an accuracy $\pm0.023$, implying that the correction is required by the
data at the $5\,\sigma$ level.
}

	By contrast, the best fit values for $t_-$ are 83, 40, and 87
days for N III], N IV], and N V respectively.  See Figure \one.  The N III]
and N IV] values are statistically inconsistent at the $5\,\sigma$ level.
I will discuss the possible sources of this inconsistency in \S 5.  For
the moment, I will simply ignore it and find the best overall fit.  Using
the above described ``$2\,\sigma$ interval'' method, I find
$t_-= 75.0\pm 2.6\,$days.

\FIG\four{N IV] light curve. Similar to as Fig.\ \three\ with
$A=53$, $t_\rise=1\,$day, $\tau=170\,$days, and $Q=0.88  \pm 0.033$.
}
\FIG\five{N V light curve. Similar to as Fig.\ \three\ with
$A=55$, $t_\rise=1\,$day, $\tau=396\,$days, and $Q=0.94  \pm 0.035$.
}
	Figures \three, \four, and \five\
show the data points for N III], N IV], and N V,
together with best fit curve having $t_-=75.0\,$days and $t_+=390.0\,$days.
The best fits includes the best values of $Q$ for
adjusting the VILSPA points, and the values of these points (crosses) have
been multiplied by $Q$ before being plotted.

\chapter{Distance to SN 1987A, the LMC, and M31}

	In the previous section I derived new values for the caustic times
$t_\pm$,
$$t_- = 75.0\pm 2.6\,{\rm days},\qquad t_- = 390.0\pm 1.8\,{\rm days}.
\eqn\tpm$$
Plait et al.\ (1994) have made a new measurement of the angular major and minor
axes, $\theta_\pm$:
$$\theta_- = 1.\htp ''242\pm 0.\htp ''022\qquad\theta_+ = 1.\htp ''716\pm
0.\htp ''022.\eqn\thpm$$
If the ring is assumed to be circular, these two sets of
measurements lead respectively to two estimates for the ring's angle of
inclination, $i$, (see Paper I)
$$
i_t = 42^\circ \htp .6\pm 1^\circ \htp .1
\qquad
i_\theta = 43^\circ \htp .6\pm 1^\circ \htp .3.
\eqn\incs$$
As I showed in Paper I, the fact that these estimates are consistent with
each other implies the distance to the ring can be calculated assuming
that the ring is circular.  If the ring does turn out to be elliptical, one
must adjust the distance estimate downward by $0.4e^4$, where $e$ is the
eccentricity.  This is most likely an extremely small correction.  Using
the procedure I described in Paper I, I therefore find a distance  and
distance modulus to the supernova of
$$\dsn = 46.77\pm 0.76\,\kpc,\qquad \mu_\sn = 18.350\pm 0.035.\eqn\dsnmu$$
As Jacoby et al.\ (1992) have pointed out, if SN 1987A (and 30 Dor) lie
in the plane of the LMC, then they are $\sim 500\,$ pc closer than the
center of the LMC.  This leads to an estimate for the distance to the LMC,
$$\dlmc = 47.3 \pm 0.8\,\kpc,\qquad \mu_\lmc=18.37\pm 0.04.\eqn\dlmcmu$$
(If one assumes that the large amount of material found 300 pc in front of
SN 1987A is the true plane of the LMC, then this estimate would be
reduced by 300 pc.)\ \

	The estimate \dlmcmu\ of $\dlmc$ allows one to make an
almost equally precise estimate of the distance to M31.  The distance
modulus difference between the LMC and M31 has previously been known to
much better accuracy than the distance modulus of either.  This is because
the apparent magnitudes of RR Lyraes and Cepheids in each galaxy are much
better measured than their absolute magnitudes.  I have shown that RR Lyraes
give $\mu_{\rm M31}-\mu_\lmc =  6.04\pm 0.07$ and Cepheids give
$\mu_{\rm M31}-\mu_\lmc =  6.07\pm 0.05$ (Gould 1994b).
Combining these values with equation  \dlmcmu\ yields
$$\mu_{\rm M31} = 24.43\pm 0.06.\eqn\dmto$$

\chapter{Discussion}

	The most important shortcoming of the measurements of $t_\pm$
described in \S\ 3, and hence of the distance determinations given in \S\ 4,
is that the estimates of $t_-$ from N III] and N IV] are inconsistent.
Formally, a discrepancy exists at the $5\,\sigma$ level, so the problem
is not likely to be statistical in nature.  Moreover, from Figures \three\ and
\four, it is clear that the N III] flux initially
rises much more slowly than the overall
best fit curve, while the N IV] flux rises much more rapidly.

	One possible explanation for the discrepancy between N III] and N IV]
is that highly ionized nitrogen recombines and cascades down to N III over
the course of the first month or two after the UV flash.  In this picture,
there would be no major buildup of N IV, since it would be just one step in
the cascade.  Such a scenario is possible, depending on the details of the
physical conditions of the gas.  This explanation is appealing because it
would seem to explain the slow initial rise of N III] between $\sim 75$
and $\sim 125$ days, a period when N IV] is rising very steeply.
See Figures \three\ and \four.  However, this explanation
would imply that the second caustic in N III] should be delayed
by a similar length of time relative to N IV].  But it is clear from Figure
\two\ that the N IV] peak is approximately coincident with N III]
and, if anything, is slightly delayed relative to it.

Another plausible
explanation for the discrepancy between N III] and N IV] is that in the
nearest portions of the
ring, a larger fraction of the N III was ionized into N IV by the UV blast,
presumably because of different local conditions.  In this case, it would
be best to take a weighted average between the fits to the two light curves,
which is the approach that I have adopted.

	However, whatever the precise reason for the discrepancy,
one might also plausibly argue that the lack of agreement
between N III] and N IV] shows that the simple models that I have adopted
for the light curves are inadequate to describe them properly.  These
inadequacies do not affect the determination of $t_+$ because the time
period around the second caustic is well sampled and indeed (as noted
in Paper I), the second caustic at $t_+\sim 390\,$days is clearly visible
even in the raw data.  However, the light curve is not well sampled around
the first caustic, so that its position must be inferred from the rate
of rise at later times.  More accurate modeling might therefore be
required to estimate $t_-$ than $t_+$.  It is therefore prudent to ask
how the results are affected if all information about $t_-$ is discarded.
Assuming that the ring is circular, I then find that
$$\dsn = 46.57\pm 0.96\,\kpc\qquad ({\rm ignoring}\ t_-),\eqn\ignore$$
Relative to equation \dsnmu, the best fit changes by $<0.5\%$ and the
error increases by $<30\%$.  That is, the measurement of $t_-$ does not
significantly change the best fit nor does it substantially raise the
confidence in its value.  The importance of measuring $t_-$ is that
it confirms that the ring is circular (or that if the ring is mildly
eccentric, $e<0.4$, the distance estimate is not significantly changed).
Thus, if one adopts the viewpoint that the light curves are not well enough
modeled to measure $t_-$,
one needs some other argument to close the loophole that the ring might
be elliptical.

	It will be possible to test whether the ring is indeed circular
when the supernova shock hits the ring early in the next century.  If
the blast is circularly symmetric in the plane of the ring, then the
shock should hit the ring all at the same time.  From the arrival
times of the light from this event at different parts of the ring,
it will be possible to reconstruct the ring geometry.  In particular,
for a circular ring, the apparent incidence of the shock will move
parallel to the major axis at a constant rate.  Of course, since
the ring is lumpy and the shock is moving much slower than the speed of light,
the actual course of the shock through the ring will take several hundred
days and will be rather complicated.  Nevertheless, by comparing this
course with detailed {\it Hubble Space Telescope (HST)} images presently
available, it should still be possible to reconstruct the ring geometry.

	There are two other minor points which should also be noted.
First, my ``$2\,\sigma$ interval'' method of determining the value and
errors of $t_\pm$ may seem rather ad hoc.  I have repeated the entire
calculation using a $1\,\sigma$ interval and find that the distance
estimate changes by $<0.5\%$.  Second, while the differences between
VILSPA and GFSC normalizations are clearly significant, one may well
question whether the single-parameter adjustment that I used is adequate.
I have therefore repeated the entire calculation allowing for both
a zero-point correction and a normalization factor for the VILSPA data.
Again, I find a change in the final result $<0.5\%$.

	I should point out that I have implicitly assumed that the
ionized nitrogen emission comes from the same region as the [O III],
H$\beta$, and [N II] emission used by Plait et al.\ (1994) to measure
the angular size of the ring.  (These three elements have very similar
emission patterns to one another.)\ \
P.\ Lundqvist (private communiction, 1994)
has pointed out that N V could come from regions at smaller radii.  If so,
one might be led to underestimate the size of the ring and hence the distance
to the supernova by tracking N V.  However, N III and O III have similar
ionization potentials so that it is unlikely that the N III] curve (for
which $t_+$ is very well constrained) would be affected by such a bias.

	Finally, I consider the effect of relaxing the assumption
that the fluorescence response is prompt, whether due to recombination
or some other cause.  In any event, the general effect of a delayed response
will be to cause the light curve caustics to be delayed relative to what
they would be with prompt fluorescence.  Hence, if the light curves are
analyzed using a prompt-response expression such as equation \fluor--\convolv,
then $t_+$ and $t_-$ will be overestimated.  This in turn will lead to an
overestimate of $\dsn$.  Hence if one allows for delayed response, equations
\dsnmu, \dlmcmu, and \dmto\ become upper limits rather than identities.

{\bf Acknowledgements:}  I would like to thank P.\ Lundqvist, G.\ Newsom, and
P.\ Plait for making some very useful comments and suggestions.

\APPENDIX{A}{Data Acquisition}

	The principal obstacle to reanalyzing the data is that the only
form in which most of them are published is in $\rm 5 cm\times 5 cm$
diagrams each containing $> 150$ points (\P 1991).  Since the points are
$\sim 40\,$days in diameter with many lying on top of one another, it is
not at all easy to recover their coordinates.  I used a photocopier to enlarge
the figures to $\rm 15 cm\times 15 cm$ and projected them on graph paper.
I then measured the graph coordinates and used the projected axes to
establish a linear transformation
to the day-intensity coordinates of the figures.  I performed an initial check
on the accuracy of my results by making a MONGO plot of the calculated
point positions and superimposing this on an enlarged photocopy
transparency of the original figures.  Transcription errors as small as two
or three days were uncovered by this procedure.  I then checked the overall
accuracy in two ways.  First I compared the dates at which I had estimated
the measurements of different ions.  Since these measurements come from the
same spectrum, they should be made on the same day.  I determined that the
rms scatter in my measurements is $<2\,$days.  Next I compared the dates
that I had found with those listed in Fransson et al.\ (1989) which contains
a subset of the data.  I found that the mean error of my measurements is
$\ll 1\,$day.  These errors are much too small to affect the conclusions
of this work.  The errors
in the intensity measurements are $<1\%$ of the scatter in these measurements
and hence are completely negligible.

\bigskip
\Ref\Dwek{Dwek, E.\ \& Felten, J.\ E.\ ApJ, 1992, 387, 551}
\Ref\Fran{Fransson, C., Cassatella, A., Gilmozzi, R., Kirshner, R.\ P.,
Panagia, N., Sonneborn, G., \& Wamsteker, W.\ 1989, ApJ, 336, 429}
\Ref\Goulda{Gould, A.\ 1994a, ApJ, 425, 51 (Paper I)}
\Ref\Gouldb{Gould, A.\ 1994b, ApJ, 426, 542}
\Ref\Jac{Jacoby, G.\ H., et al.\ 1992, PASP, 104, 599}
\Ref\Jak{Jakobsen, P., et al.\ 1991, ApJ, 369, L63}
\Ref\Lund{Lundqvist, P.\ 1994, 34th Herstmonceux Conference, eds.\
R.\ Clegg, P.\ Meikle, \& I.\ Stevens, in press}
\Ref\Pana{Panagia, N., Gilmozzi, R., Macchetto, F., Adorf, H.-M. \&
Kirshner, R.\ P.\ 1991, ApJ, 380, L23}
\Ref\plait{Plait, P., Lundqvist, P., Chevalier, R., \& Kirshner, R.\ 1994,
ApJ, submitted}

\refout
\endpage
\figout
\end